\documentclass[12pt]{article}
\usepackage{titlesec}
 \usepackage{multicol}
 \usepackage{graphics,natbib,graphicx,latexsym,amssymb,epsfig,bm,amsmath,fullpage,float}
 \usepackage[toc]{appendix}
 \usepackage{mathrsfs,amsfonts,amssymb}
 \usepackage{multirow,rotating,sectsty}
 \usepackage{longtable}
 \usepackage{amsthm}
 \usepackage{amsmath}
 \usepackage{graphicx}
 \usepackage{epsfig}
 \usepackage{amsthm}
 \usepackage{amsmath}
 \usepackage{amsbsy}
 \usepackage{amsfonts}
 \usepackage{enumitem}
 \usepackage[T1]{fontenc}
 \usepackage[utf8]{inputenc}
 \usepackage[english]{babel}
 \usepackage{filecontents}
 \usepackage{subcaption}
 \usepackage{threeparttable}
 \numberwithin{equation}{section}
 \usepackage[usenames]{color}

 \newtheorem*{remark}{Remark}
 \titleformat{\paragraph}
 {\normalfont\normalsize\bfseries}{\theparagraph}{1em}{}
 \titlespacing*{\paragraph}
 {0pt}{3.25ex plus 1ex minus .2ex}{1.5ex plus .2ex}

\begin{document}
\title{Efficient Data Reduction Strategies for Big Data and High-Dimensional
LASSO Regressions}

\author{Xin Wang \\
    University of Illinois-Chicago,\\ Chicago, USA (wangxin100037@gmail.com)\\
     \\
    Min Yang \\
    University of Illinois-Chicago,\\ Chicago, USA (myang2@uic.edu)\\
    \\
    William Li\\
    Shanghai Advanced Institute of Finance\\ Shanghai, China (wlli@saif.sjtu.edu.cn)
    }

\date{}
\maketitle
 	
\begin{abstract}
The IBOSS approach proposed by \cite{WangYangStufken2016} selects the most informative subset of $n$ points. It assumes that the ordinary least squares method is used and requires that the number of variables, $p$, is not large. However, in many practical problems, $p$ is very large and penalty-based model fitting methods such as LASSO is used. We study the big data problems, in which both $n$ and $p$ are large. In the first part, we focus on reduction in data points. We develop theoretical results showing that the IBOSS type of approach can be applicable to penalty-based regressions such as LASSO. In the second part, we consider the situations where $p$ is extremely large. We propose a two-step approach that involves first reducing the number of variables and then reducing the number of data points. Two separate algorithms are developed, whose performances are studied through extensive simulation studies. Compared to existing methods including well-known split-and-conquer approach, the proposed methods enjoy advantages in terms of estimation accuracy, prediction accuracy, and computation time. 

\end{abstract}

\noindent%
{\it Keywords:}   Data reduction, IBOSS, high-dimensional data, sure independence screening.

\section{Introduction}
 	
Extraordinary amounts of data are being produced daily in today's world. 
The challenge of analyzing big data can result from both the extremely large number of data points (denoted by $n$) and the very large number of variables (denoted by $p$). Consequently, the data can be both {\it big} and {\it high-dimensional}. 
For many such problems, {\it data reduction} is often a necessary first step. Various data reduction methods have been proposed in recent years. Some existing methods are not model dependent, such as uniform sampling, the clustering method of \cite{HarpeledKushal2007}, and the support point method of \cite{MakJoseph2018}. These methods have been shown to work effectively. Model-based data reduction methods have also been considered,
such as the leveraging sampling method discussed in \cite{PingMa2014-JMLR}.  While the random sampling approaches are easy to implement in practice, they have the limitations in terms of how much information can be extracted from the full data. For example, \cite{WangYangStufken2016} showed that the information contained in the subdata was bounded by the size of the subsample. This motivated \cite{WangYangStufken2016}  to propose an information-based optimal subdata selection (IBOSS), which chooses a group of most informative data points from the whole data set.
The IBOSS idea is similar to that of the exact optimal design, where a set of design points are selected from design space. For example,  \cite{ChenHuang2000} studied the construction of the exact D-optimal designs for weighted polynomial regression models. However, a data point cannot be selected more than once in subdata selection while there is no such restriction in an exact design. When the size of full data is extremely large, the exact design approach may not be feasible for a subdata selection.
The basic motivation of the IBOSS method is rooted in approximate optimal experimental design \citep{kiefer1959optimum}, where optimal weights instead of exact number of design points are considered.

The novel approach of IBOSS takes advantage of nice properties of the information matrix resulting from the ordinary least squares (OLS) method, and thus is mostly suitable for problems where $p$ is relatively small. In practice, there are many situations where $p$ is large. 
The ordinary least squares method is generally not effective in solving such problems. Instead, it has been shown repeatedly in the literature that alternative methods based on penalized regressions are more effective for variable selections for data with large $p$.
Approaches targeting high-dimensional variable selection include a series of penalized regression approaches like LASSO \citep{tibshirani1996regression}, ridge \citep{hoerl1970ridge}, elastic net \citep{zou2005regularization}, SCAD \citep{fan2001variable} and so on. There were also studies focusing on even higher dimensional cases, where $p>>n$, such as the Dantzig selector of \cite{candes2007dantzig} and the Sure Independence Screening (SIS) methods of \cite{fan2008sure}.

When $p$ is large and there is a penalty term in the model, some foundational results of the IBOSS of Wang et al. (2019) no longer hold. Consequently, some natural and important research questions arise: (i) {\it Can we use the IBOSS type of approach for the variable selection model having a penalty term?} (ii) {\it What modifications of IBOSS are needed when $p$ is extremely large?} (iii) {\it For larger data problems like this, are there better benchmark methods to compare with, besides the commonly used uniform and leverage sampling methods that were considered in almost all IBOSS related works?} In this article we aim to address these three issues for data reduction problems with large $n$ and large $p$. First, we develop an IBOSS-LASSO algorithm when $p$ is relatively large. 
An important condition in the framework of IBOSS is the existence of a direct link between the estimator and the inverse of the information matrix \citep{WangYangStufken2016}. Unfortunately, for penalty based methods like LASSO, the information matrix cannot be properly defined. We develop some theoretical results and show that the LASSO estimator is still affected by the usual information matrix in a very similar way as OLS under mild conditions. The extensive simulation studies conducted in later sections demonstrate that the IBOSS-LASSO algorithm works both effectively and efficiently.

The second main contribution of this article is the introduction of a two-step approach. A necessary condition for IBOSS type of approaches is $k \ge 2p$. For an ultra-high dimensional problem, this may not be feasible. It has been shown in the literature that sparsity (i.e., only a small subset of $p$ variables (features) are truly associated with the response) comes frequently with high-dimensional data \citep{fan2008sure}. Our two-step approach consists of the reduction in both variables and data points. In the first step, we reduce the number of variables by choosing those having the largest absolute correlations with the response, an idea first proposed in \cite{fan2008sure} and called Sure Independence Screening (SIS). Then in the second step, we reduce the number of data points by choosing the most informative data points. The combined approach is called the SIS-IBOSS algorithm, whose performances are studied in details in later sections.

The third contribution is to compare the proposed IBOSS algorithms with the split-and-conquer approach, for which the existing literature on IBOSS type of works generally missed. Unlike
{\it subdata}-based methods, the split-and-conquer approach works directly on the {\it full data}.  It randomly splits the full dataset into a few subsets, analyzes each subsets separately, and combines results from each analysis together. Utilizing a  majority voting method, \cite{chen2014split}  developed a split-and-conquer approach to analyze a dataset with large $n$ and $p$. By utilizing the parallel computation technology, the computation time often becomes affordable for dealing with the full data. In this manuscript, a systematic comparison between the two methods is implemented.

The reminder of article is organized as follows. Section 2 introduces the IBOSS-LASSO algorithm. Some theoretical justifications of the proposed algorithm are provided. Section 3 discusses the SIS-IBOSS algorithm. In Section 4 we show theoretical computation cost of methods. Extensive numerical comparisons between the proposed methods and other approaches are presented in Section 5. The performances of the proposed methods are examined through a real data example in Section 6.
Concluding remarks are given in Section 7.



\section{IBOSS-LASSO}

\subsection{Linear Model and LASSO Estimator}
 	
Denote the whole data set as $(\bm{x}_1,y_1),..., (\bm{x}_n,y_n)$ and assume the linear model:
 	\begin{equation}\label{lm}
 	E(y_i) =\bm{x}_i^T\bm{\beta},\ i = 1,2,..., n,
 	\end{equation}
where $\bm{x_i}=(1,x_{i1},..., x_{ip})^T$ and $y_i$ refer to the $i$-th point, and $\bm{\beta}=(\beta_0,\beta_1,..., \beta_p)$ is a $(p+1)$-dimensional vector of parameters. 
  	 	Let $l(\bm{\beta};\bm{y,\bm{X}})$ be the log-likelihood function. Then the penalized regression estimator in general form can be given by
 	 	\begin{equation*}
 	 	\begin{aligned}
 	 	\hat{\bm{\beta}}^{(penalized)} = \arg\max_\beta\left\{\dfrac{l(\bm{\beta},\bm{y},\bm{X})}{n}-\rho(\bm{\beta},\lambda) \right\},
 	 	\end{aligned}
 	 	\end{equation*}
 	 where	$\rho(\bm{\beta},\lambda)$ is the penalty function with the tuning parameter $\lambda$, shrinking the original maximum likelihood estimator.
 	 	Popular choices of $\rho(\bm{\beta},\lambda)$ include LASSO, ridge, elastic net, SCAD and so on.
 The tuning parameter	 	$\lambda$ is usually determined by minimizing the cross-validation error.
 	 	
 	 	 LASSO with an $L_1$ penalty term does variable selection and shrinkage simultaneously, reducing the number of variables as well as the variance of the estimator. When $\rho(\bm{\beta},\lambda) =\lambda\sum_{j=1}^p|\beta_j|$, the LASSO estimator for linear regression is
 	\begin{equation}
 	\hat{\beta}^{LASSO}= \arg\min_\beta\dfrac{1}{n}\sum_{i=1}^n(y_i-\bm{x}_i\bm{\beta})^2 + \lambda\sum_{j=1}^p|\beta_j|.
 	\end{equation}
 	This is equivalent to minimizing $\sum_{i=1}^n(y_i-\bm{x}_i\bm{\beta})^2 $  with constraint $\sum_{j=1}^p|\beta_j|<S, S>0$.


The LARS algorithm \cite{efron2004least} provides an efficient algorithm for implementing LASSO. Another commonly used algorithm is the coordinate descent algorithm, which was proposed by Friedman et al. (2007) 
to address large data LASSO problems with sparsity. This algorithm iteratively optimizes a target function (the sum of likelihood and penalty) on one dimension at a time. It provides a fast and robust solution to LASSO and other penalized regressions. In this article we use the coordinate descent algorithm to implement LASSO regressions.
 	
  	\subsection{Challenges in Using IBOSS for LASSO}\label{theory}
 	
The IBOSS approach has been proved to be an effective and efficient algorithm fitting linear models using the least squares method Wang et al. (2019).
In traditional studies of optimal design, the Fisher information matrix could be written as a negative Hessian matrix of a log likelihood function.
Let $\delta_i=1$ if the $i$th data point is in the subdata and $\delta_i=0$ otherwise. We want to select $\bm{\delta}=(\delta_1,\ldots,\delta_n)$ subject to $\sum_{i=1}^n\delta_i=k$, such that it ``maximizes'' the information matrix. For the linear model (\ref{lm}), the information matrix is
 	 		\begin{equation}
 	 		\bm{M}({\bm{\delta}}) =\dfrac{1}{\sigma^2}\sum_{i=1}^n \delta_i\bm{x_ix_i}^T.
 	 		\end{equation}
It is well known that the covariance matrix $V(\hat{\bm{\beta}})$ is directly proportional to the inverse of the information matrix. From the perspective of data reduction, an ``optimal'' subset of $k$ points $\bm{\delta}^*$ would maximize a function, say, the determinant of the information matrix $\bm{M}(\bm{\delta})$. This is the $D$-optimal subset that satisfied $\delta^*=argmax|\bm{M}(\bm{\delta})|$ over all possible size-$k$ subsets $\delta$. When $n$ is large, the total number of possible subsets, which equals $(^n_k)$, can be prohibitively large. 
 	
The main challenge in using IBOSS for LASSO regressions is that, when penalty terms are involved, the above desired property of the information matrix no longer holds for two reasons. First, the information matrix is now derived as the negative expectation of the second derivative of the log-likelihood function, which may not be properly defined for the $L_1$ penalty. Second, a penalty term is added to the optimization function as a separate part. Even if the second derivative of the penalized likelihood function could be defined, its direct connection to estimator variance would not hold like that for MLE.
 	
In the next subsection we show how to overcome these challenges from the theoretical point of view. More specifically, we will demonstrate that, although the variance of the LASSO estimator of the selected subdata through $\bm{\delta}$ 
is not proportional to $M(\bm{\delta})^{-1}$, it is still affected by $M(\bm{\delta})$ in a way similar to OLS under mild conditions.
Thus, the IBOSS approach, though intended to optimize the information matrix of OLS estimator, is also applicable and efficient for the LASSO estimator.
 	
\subsection{LASSO Asymptotics and Connection with Information Matrix}\label{lassoprop}

For a design matrix $\bm{X} = (\bm{x}_1,...,\bm{x}_n)^T$, assume
the following  regularity conditions: (i)
  	$\dfrac{1}{n}\max_{1\le i\le n}\bm{x_i}^T\bm{x_i} \to 0,$  	
 and (ii)	
 	$\bm{C}_n = \dfrac{1}{n}\sum_{i=1}^n\bm{x_ix_i}^T \to \bm{C}$,
 	where $C$ is a nonnegative definite matrix. (Note that $\bm{C}_n$ is proportional to the information matrix for using the full data $\bm{M} =\dfrac{1}{\sigma^2}\sum_{i=1}^n \bm{x_ix_i}^T$.)
\cite{knight2000asymptotics} proved the following asymptotic property of LASSO-type estimators:

 		If $\lambda_n/\sqrt{n}\to \lambda_0\ge 0$ and $C$ is nonsingular then
 		\begin{equation}\label{consi}
 		\sqrt{n}(\hat{\bm{\beta}}_n^{LASSO}-\bm{\beta}) \to_d \arg\min_{\bm{u}}V(\bm{u}),
 		\end{equation}
 		where
 		\begin{equation}\label{vu}
 		V(\bm{u}) = -2\bm{u}^T\bm{W}+\bm{u}^T\bm{C}\bm{u} + \lambda_0\sum_{j=1}^p\left[u_j\text{sign}(\beta_j)I(\beta_j\ne 0)+|u_j|I(\beta_j=0)\right]
 		\end{equation}
 		and $\bm{W}\sim N(\bm{0},\sigma^2\bm{C})$.

While (\ref{consi}) gives the asymptotic distribution of $\hat{\bm{\beta}}_n^{LASSO}$, the explicit expression of $\arg\min_{\bm{u}}V(\bm{u})$, which is critical for a criterion for selecting a subdata, is not available due to the last term of (\ref{vu}). However, if we can assume that the tuning parameter $\lambda_n$ grows slower than $\sqrt{n}$, (i.e.,  $\lambda_n = o(\sqrt{n})$, and $\lambda_n/\sqrt{n}\to \lambda_0=0$,) then
$ \lambda_0\sum_{j=1}^p\left[u_j\text{sign}(\beta_j)I(\beta_j\ne 0)+|u_j|I(\beta_j=0)\right] = 0$.  In this case, we have:
 	\begin{equation}\label{ntoinfty}
 	\begin{aligned}
 	V(\bm{u}) &=  -2\bm{u}^T\bm{W}+\bm{u}^T\bm{C}\bm{u},\\
 	\arg\min_{\bm{u}}V(\bm{u}) &= \bm{C}^{-1}\bm{W},\\
\text{ and } 	\sqrt{n}(\hat{\bm{\beta}}_n^{LASSO} - \bm{\beta} )&\to_d \bm{C}^{-1}\bm{W} \sim N(0,\sigma^2\bm{C}^{-1}).
 	\end{aligned}
 	\end{equation}
Under this condition, the asymptotic behavior of $\hat{\bm{\beta}}_n^{LASSO}$ is as the same as that of OLS estimator.

In practice, requiring that the tuning parameter $\lambda_n$ grows slower than $\sqrt{n}$ can be considered as a mild condition. As shown in simulation studies given in this article, even if this condition is not met, the subdata selected through the criterion based on (\ref{ntoinfty}) is still more informative than that of random sampling.

From (\ref{ntoinfty}) we can conclude that as long as $\lambda_n =o(\sqrt{n})$, $\hat{\bm{\beta}}^{LASSO}_n$ is a consistent estimator and its estimation variance is asymptotically proportional to $\bm{C}^{-1}$, which is the limit of the inverse of information matrix. This conclusion builds a connection between the variance matrix of the LASSO estimator and the information matrix when $n\to\infty$. Consequently, a subset $\delta$ maximizing $|M(\bm{\delta})|$ can be considered as ``optimal'' in an asymptotic way. This motives us to present an IBOSS-type of approach when LASSO is used for model fitting.


\subsection{IBOSS-LASSO Algorithm}\label{IBOSS_Alg}

We now formally describe the IBOSS-LASSO algorithm as below:

 	Algorithm 1. ($D$-optimal IBOSS), assume $r =  k/2p$ is an integer.
 	
\begin{enumerate}
 	\item Start from $x_{i1}$,$1\le i\le n$. In the full data pool, select $r$ points with the smallest $x_{i1}$ and $r$ with the largest $x_{i1}$ values. Update the pool of data by excluding selected points.

 	\item For $j=2,\ldots,p$, select $r$ points with the smallest $x_{ij}$ and $r$ points with the largest $x_{ij}$ values from the pool of data. Update the pool by excluding selected points.

 		\item After $p$ iterations, we obtain the $D$-optimal subdata $\bm{\delta}_D$. Compute the LASSO estimator $\hat{\bm{\beta}}^{LASSO}_D$ from the subdata, where $\hat{\bm{\beta}}^{LASSO}_D =  \arg\min_\beta\sum_{i\in \bm{\delta_D}}(y_i-\bm{x}_i\bm{\beta})^2 + \lambda_{D}\sum_{j=1}^p|\beta_j|$. The tuning parameter $\lambda_D$ is chosen by using 10-fold cross-validation on the subdata $\bm{\delta}_D$.
 	\end{enumerate}
 	\noindent\hrulefill
 	
The theoretical result in Section \ref{lassoprop} showed that the subdata chosen by this algorithm contain maximum amount of information as long as $\lambda_n/\sqrt{n}\to 0$. In later sections of this article, we will conduct thorough simulation studies, which demonstrate that IBOSS-LASSO outperforms competing approaches in terms of both statistical efficiency and computational complexity. It also provides empirical evidence that, even if $\lambda_n/\sqrt{n}\to  0$ does not hold, the algorithm can still perform well.
 	 	
Notice that the IBOSS approach is also suitable for parallel computing since the partial sorting is done separately on each column of the dataset. One can simultaneously process each column and combine the indices to form the subdata $\bm{\delta}_D$.

\section{SIS-IBOSS approach}\label{sis}

For data with a very large number of variables, Fan and Lv (2008) proposed a sure independence screening approach (SIS). They sorted the variables by their sample correlations with $\bm{y}$ and define a sub-model $\mathscr{M}_\gamma$ with only the first $\tilde{p}$ variables ($\tilde{p}=[\gamma p] <p$), where $\gamma$ denotes the proportion of the variables that are included for model fitting.
Under the setting of $n<<p$, Fan and Lv (2008) identified conditions under which the sure screening property
\begin{equation}
\label{SIS}
\lim_{n\rightarrow \infty}P(\mathscr{M}_*\subset\mathscr{M_\gamma}) = 1
\end{equation}
holds for LASSO problems, where $\mathscr{M}_*$  is the true model. Relation (\ref{SIS}) shows that the sub-model consisting of variables that are highly correlated with the response has a larger probability to contain the true model for larger $n$.

Inspired by the SIS approach, we propose a two-step algorithm called SIS-IBOSS. In the first step, the algorithm selects a subset of $\tilde{p} < p$ variables that have the largest absolute correlations with the response. Then it uses an IBOSS type of approach by choosing extreme points with respect to each of the $\tilde{p}$ variables. Reducing the number of variables at the first step would not only make IBOSS applicable, it could also improve the performance of IBOSS. Given the sparse assumption of LASSO model, selecting observations with extreme values in false variables $\bm{X}_j$ may not be helpful for reducing the variance of $\hat{\bm{\beta}}^{LASSO}$, because $\hat{\beta}_j^{LASSO}$ is likely forced to be 0 under $L_1$ penalty. In situations like this, the information matrix should be based on the variables excluding those false variables. It is likely more efficient if we only consider variables with non-zero coefficients when we select the subdata.
 	
%
%
The two-step SIS-IBOSS Algorithm is given as as follows:

 	\noindent\hrulefill

Algorithm 2. (SIS-IBOSS). Assume $r =  \dfrac{k}{2\tilde{p}}$ is an integer, where $0<\tilde{p}\le p$ is the number of variables selected in Step 1.
 	
\begin{enumerate}
\item Variable reduction using SIS: Calculate $corr(\bm{X_{j}},\bm{y})$, $j=1,...p$, and select $X_{(1)},...X_{(\tilde{p})}$ with the largest $|corr(\bm{X_{j}},\bm{y})|$, where $X_j$ is $(j+1)$-th column of $\bm{X}$.

\item Data reduction using IBOSS: Start from $x_{i(1)}$,$1\le i\le n$. In the full data pool, select $r$ points with the smallest $x_{i(1)}$ and $r$ points with the largest $x_{i(1)}$ values. Update the pool of data by excluding selected points.

    For $j=2,...\tilde{p}$, select $r$ points with the smallest $x_{i(j)}$ and $r$ points with the largest $x_{i(j)}$ values from the pool of data. Update the pool by excluding selected points.

\item Fitting model using LASSO: Denote the $D$-optimal subdata after $\tilde{p}$ iterations as $\bm{\delta}_{SD}$. Compute the LASSO estimator $\hat{\bm{\beta}}^{LASSO}_{SD} =  \arg\min_\beta\sum_{i\in \bm{\delta_{SD}}}(y_i-\bm{x}_i\bm{\beta})^2 + \lambda_{SD}\sum_{j=1}^p|\beta_j|$ from the subdata. Here, the tuning parameter $\lambda_{SD}$ is chosen by 10-fold cross-validation on the subdata $\bm{\delta}_{SD}$.
 	\end{enumerate}
 	\noindent\hrulefill


\begin{remark}
 	SIS-IBOSS approach takes advantage of both SIS and IBOSS approaches. The SIS approach identifies the false variables using all observations with relative cheap computation cost. Notice that the information matrix is based on the variables excluding those false variables. Consequently, SIS-IBOSS approach could reduce the variance of $\beta_j^{LASSO}$ more efficiently.
 \end{remark}
 	

\section{Computational Complexity}
 	
\subsection{IBOSS-LASSO Computational Complexity}\label{comp_theo}
In all numeric studies of this article, LASSO is implemented using the coordinate descent algorithm. \cite{wu2008coordinate} and \cite{friedman2010regularization} both demonstrated that this algorithm was ``considerably faster and more robust'' compared to LARS on $L_1$ and $L_2$ penalized regressions.  At each iteration, the algorithm requires $O(n)$ operations to update each coordinate, $O(np)$ operations in total to update $p$ coordinates. Denoting the number of iterations before convergence by $R_{iter}$, the computational complexity for using LASSO on a subset of size $k$ equals $O(kpR_{iter})$, much smaller compared to $O(npR_{iter})$ for using the full data. The savings are even more substantial when compared to the complexity of $O(p^3+np^2)$ for the LARS algorithm \cite{efron2004least}.
 	
Unlike the simple random sampling approach, both algorithms proposed in this article require an extra step of sorting columns. For each variable, the algorithms select the top $r$ and bottom $r$ elements after sorting. Using a partial sorting algorithm with average computational complexity of $O(n+r\log (r))$ \citep{martinez2004partial}, it can be easily seen that the complexity for sorting amounts to $O(np+pr\log(r))$ and $O(n\tilde{p}+\tilde{p} r\log(r))$ for IBOSS-LASSO and SIS-IBOSS, respectively.
 	
In total, the computational complexity is $O(np+k\log(r)+kpR_{iter})$ for IBOSS-LASSO and $O(\gamma np+\gamma kpR_{iter}+k\log(r))$  for SIS-IBOSS, both of which are considerably smaller compared to $O(npR_{iter})$ for the full-data approach. Furthermore, the difference between computation times of the two approaches is amplified when cross-validation is used, in which we need to search for the best tuning parameters.
 	
 	\subsection{Subset Approaches}
The existing  subset selection methods are mainly based on simple random sampling or leveraging  sampling. The latter one uses the statistical leverage score,  $\bm{x}_i^T(\bm{X}^T\bm{X})^{-1}\bm{x}_i$ as the probability of including point $i$ in the sample. The complexity for computing all $n$ leverages scores is $O(np^2)$. An alternative and faster approach is to use a randomized algorithm to approximate leverage scores (Drineas et al. 2012), whose computational complexity is given by  $O(np\log n/\epsilon^2)$, where $\epsilon \in(0,0.5]$.
Although the leveraging method is not specifically designed for LASSO, we include this approach when we evaluate the performances of different subset selection methods.


\subsection{Split-and-Conquer Approach}
Besides subdata-based methods, another important way to save computation time is parallel computation.
A split-and-conquer approach \cite{chen2014split} randomly separates the whole dataset into $K$ subsets $\bm{X}_1,...\bm{X}_K$, fits penalized regressions separately, and obtains estimators $\hat{\bm{\beta}}^{(1)},\ldots,\hat{\bm{\beta}}^{(K)}$. \cite{chen2014split} proposed to use a simple majority voting method to determine the number of non-zero coefficients, in which $\beta_j^{(c)} \ne 0$ if $\sum_{k=1}^KI(\beta_j^{(k)}\ne 0)>w$, where $w$ is a pre-specified threshold value. Then they used a weighted linear combination formulate to combine $K$ estimators:
\begin{equation}
\hat{\bm{\beta}}^{(c)} = \bm{A}\left(\sum_{k=1}^K\bm{A}^T(\bm{X}_k^T\bm{X}_k)\bm{A}\right)^{-1}
\sum_{k=1}^K\bm{A}^T\bm{X}_k^T\bm{X}_k\bm{A}\hat{\bm{\beta}}^{(k)}_{\hat{A}_c},
\end{equation}
where $\hat{A}_c = \{j:\hat{\beta}_j^{(c)}\ne 0\}$ and $A$ is formed by the columns of $p\times p$ identity matrix whose indices are in $\hat{A}_c$.


The main advantage of this approach is that it takes advantage of the power of parallel computing. Under certain conditions, the estimator is proved to be asymptotically equivalent to the one obtained from analyzing the entire data. Using LARS to fit LASSO regression, \cite{chen2014split} showed that the split-and-conquer approach can save up to $(1-1/K^{(a-1)})\%$ computing time. We compare the proposed IBOSS type of algorithms with the split-and-conquer method in the next section. For fair comparisons, we also use the coordinate descent algorithm for the split-and-conquer approach.

Compared to IBOSS, the split-and-conquer approach has an extra step in aggregating estimators in the final step, which involves computing the inverse of a large matrix. This step typically takes much more time compared to the partial sorting step of IBOSS. We shall investigate the computation time difference through extensive simulation studies in Section \ref{numeric}.

\section{Numerical Studies}\label{numeric}

 	\subsection{Simulation Settings}

	Data are generated from a linear model $y_i=\bm{x}_i\bm{\beta}+\epsilon_i$, where $\epsilon_i$'s ($i=1,...,n$) are I.I.D. $N(0,1)$ errors. The performances of different methods are evaluated over a variety of scenarios with respect to different full sample sizes, number of variables, and subsample sizes. In a series of tables and figures, the following symbols are used:
 		\begin{itemize}
			\item Full: LASSO regression is performed on the full data.
			\item SPC(w/K): LASSO regression is performed on $K$ subsets of data. To obtain the combined estimator of $\bm{\beta}$, $\beta_j$ is set to be a weighted average from all the $K$ estimation only when at least $w$ of $K$ votes are non-zero, otherwise $\beta_j$ is set to be 0.
			\item D-OPT(IBOSS-LASSO): LASSO regression is performed on the subset of data selected by IBOSS approach, as shown in Algorithm 1.
 			\item SIS-IBOSS($s$): SIS is first used to select the $s$ variables having the largest absolute values of correlation with $\bm{y}$. Then the LASSO regression is used, as detailed in Algorithm 2.
 			\item LEV: LASSO regression is performed on the subset of data selected by leverage sampling approach, where exact leveraging scores are calculated with all variables.
 			\item LEV($s$): LASSO regression is performed on the subset of data selected by leveraging sampling approach, where leveraging scores are calculated with only $s$ variables with largest absolute values of correlation coefficient  with $\bm{y}$.
 			\item ALEV($s$): It is similar to LEV($s$) except that leveraging scores are computed  by an approximate algorithm by Drineas and et al. (2012).
 			\item UNIF: LASSO regression is performed on simple random sample of data.
 		\end{itemize}
 		
In all scenarios described above, several distributions are used to generate the design matrix $\bm{X}$ (except $x_{i0}=1$):
 		 \begin{itemize}
 		 	\item $x_{ij}\sim N(0,1), j=1,...,p $, standard normal distribution.
 		 	\item  $x_{ij} \sim \log(N(0,1)),j=1,...,p$, log standard normal distribution.
 		 	\item $x_{ij} \sim t(df=2),j=1,...,p$, t-distribution with two degree of freedom.
 		 	\item $x_{ij} = 0.25Z_1+0.25Z_2+0.25Z_3+0.25Z_4 $, where
 		 	$Z_1\sim N(0,1)  $,
 		 	$Z_2\sim t(2) $,
 		 	$Z_3\sim t(3)$,
 		 	$Z_4\sim \log(N(0,1))$, a mixed distribution.
 		 \end{itemize}

In each simulation setting, the true model and the full data set remain the same across all methods. For methods using subdata, the size of the subdata also remains the same, although the actual subset of data may vary depending on the selecting method.
Each true model contains $p^*=\lfloor\sqrt{p}\rfloor+1$ coefficients that are generated from a normal distribution with mean of $\beta^*=\sqrt{\frac{\log(5000)}{1000}}/2$ and the standard deviation of $\beta^*/5$, and all other $p-p^*$ coefficients are set as 0.
 	 	 	
We evaluated performances of both IBOSS-LASSO (when $k \ge 2p$) and SIS-IBOSS (when $k < 2p$). All simulation studies are conducted with 100 replications, and the results reported in the following subsection are calculated by averaging over the $100$ replications. We used the $glmnet$ function in R package to do LASSO regressions. To find the best tuning parameter $\lambda$, $10$-fold cross-validation was used. More specifically, $\lambda$'s are generated by the following scheme \cite{friedman2010regularization}: Start from the smallest value $\lambda_{max}$ for which entire $\hat{\bm{\beta}}=0$. Select a minimum value $\lambda_{min}=\epsilon\lambda_{max}$ and construct a sequence of $C$ values of $\lambda$ decreasing from $\lambda_{max}$ to $\lambda_{min}$ on the log scale. In all studies, we used typical values of $\epsilon=0.001$ and $C=100$.

 	\subsection{Computation Time}

\begin{table}[tp]
\centering
\caption{CPU time (second) for various $n$, $p$, $k$($p<k$), with $X\sim t(2)$}
\label{table1}
\begin{subtable}{.65\textwidth}
\centering
\caption{Different $p$ with $n = 100,000, k=1,000$}\label{tb1}
\begin{tabular}{rr|rrrr}
\hline
$p$&$p^*$&Full  &LEV    &D-OPT        &UNIF\\
\hline
50 &  8&  3.9838&  1.2283& 0.3266& 0.1267\\
100& 11& 10.1047&  2.4081& 0.6031& 0.1968\\
500& 23&113.5003& 26.5367& 3.2721& 3.3062\\ \hline
\end{tabular}
\end{subtable}
\begin{subtable}{.65\textwidth}
\centering
\caption{Different $n$ with $p = 500,k=1,000,p^* =23$ }\label{tb2}
\begin{tabular}{r|rrrrr}
\hline
$n$   &Full    &LEV       &D-OPT        &UNIF\\
\hline
10,000 & 15.6441& 5.6652& 3.1005& 3.4587\\
20,000 & 27.7483& 7.7508& 2.7781& 3.3517\\
40,000 & 51.2922&12.6871& 2.8129& 3.3684\\
80,000 & 90.8922&21.4146& 2.9775& 3.1412\\
100,000 &113.5003&26.5367& 3.2721& 3.3062\\ \hline
\end{tabular}
\end{subtable}
\begin{subtable}{.65\textwidth}
\centering
\caption{Different $k$ with $p = 500,n = 100,000,p^*=23$ }\label{tb3}
\begin{tabular}{r|rrrrr}
\hline
$k$   &Full    &LEV       &D-OPT  &UNIF\\
\hline
1,000&113.5003& 26.5367& 3.2721& 3.3062\\
2,000&113.7291& 27.4662& 4.3644& 4.2358\\
3,000&113.6444& 28.6134& 5.5733& 5.7411\\
4,000&113.3549& 29.9633& 6.8874& 7.2475\\
5,000&115.0434& 31.8262& 8.2643& 8.7430\\ \hline
\end{tabular}
\end{subtable}
\end{table}
 	 	
Table \ref{table1} compares CPU time of the model fitting approaches using (1) the full data (denoted as Full), (2) subdata chosen from leverage sampling (denoted as LEV), (3) subdata chosen from IBOSS-LASSO (denoted as D-OPT), and (4) subdata chosen from uniform sampling (denoted as UNIF). All computations are carried out on a Dell workstation with a dual Intel Xeon processor and 128 GB memory. The computation time unit is second. Not surprisingly, the IBOSS-LASSO approach results in substantial savings in computation time compared to model fitting using the full model. The proposed IBOSS algorithm is also much faster than the leverage sampling method. On average, the former method took about 1/4 of the CPU time of the latter.

It is interesting to note that the uniform sampling method, which is the simplest one to implement, was not necessarily the fastest approach. Table \ref{tb1} shows that for cases of small $p$, it was faster than the corresponding IBOSS method. However, when $p$ is as large as 500, the IBOSS-LASSO method was the fastest approach. To further explore the reason behind this phenomenon, Table \ref{table2} shows the split CPU times of model fitting using subdata: (1) the time for subdata selection, and (2) the time for model fitting using LASSO. Table \ref{tb3b} shows clearly that the model fitting time using LASSO was the smallest when the subdata was chosen from the IBOSS approach in all cases. This provides strong empirical evidence that points selected from IBOSS are indeed very informative and facilitate the model selection procedure of LASSO.


\begin{table}[thpb]
\caption{CPU time (second) by steps with fixed $p = 500,p^*=23,X\sim t(2),$}
\label{table2}
\centering
\begin{subtable}{.8\textwidth}
\centering
\caption{Subset Selection $k=1,000$ from various $n$}\label{tb3a}
\begin{tabular}{r|rrrrr}
\hline
$n$ &Full    &LEV       &D-OPT  &UNIF\\ \hline
10,000&0&  2.5834& 0.6837&  5e-04\\
20,000&0&  4.9825& 0.7918&  6e-04\\
40,000&0& 10.1231& 1.0725&  7e-04\\
80,000&0& 19.2631& 1.5572&  8e-04\\
100,000&0& 24.4773& 1.8986&  7e-04\\ \hline

\end{tabular}
\end{subtable}
\begin{subtable}{.8\textwidth}
\centering
\caption{LASSO Regression $k=1,000$ from various $n$}\label{tb3b}
\begin{tabular}{r|rrrrr}
\hline
$n$ &Full    &LEV       &D-OPT &UNIF\\ \hline
10,000& 15.6441& 3.0818 &2.4168 & 3.4582\\
20,000& 27.7483& 2.7683 &1.9863 & 3.3511\\
40,000& 51.2922& 2.5640 &1.7404 & 3.3677\\
80,000& 90.8922& 2.1515 &1.4203 & 3.1404\\
100,000&113.5003& 2.0594 &1.3735 & 3.3055\\ \hline
\end{tabular}
\end{subtable}
\end{table}


 	\begin{table}[htpb]
 		 	 	\centering
 		 	 	\caption{CPU time (second) for different settings of $n$, $k$ with a fixed $p=5,000$ ($k \le p$).}
 \label{table3}
 		 	 	 	  	\begin{subtable}{.7\textwidth}
 		 	 	 	  		\centering
 		 	 	 	  		\caption{Various $n$ with $p = 5,000,p^*=71, k=1,000,X\sim t(2)$}\label{klep2}
\begin{tabular}{r|rrrrr} \hline
$n$ &Full &LEV($250$)&ALEV($250$)&SIS-IBOSS($250$)&UNIF\\ 	\hline
10,000 &  237.5579& 10.8103&  10.3443& 10.0679& 9.7813\\
20,000&  328.0933& 12.0154&  10.8449& 10.2062& 9.6441\\
40,000&  519.4563& 14.0740&  11.3624& 10.4938& 8.9245\\
80,000&  912.8601& 18.6296&  13.4083& 11.7900& 7.9334\\
100,000& 1132.2454& 21.5109&  15.4830& 13.2781& 8.0193\\ \hline
 		 	 	 	  		\end{tabular}
 		 	 	 	  	\end{subtable}
 	 \begin{subtable}{.7\textwidth}
 	 	\centering
 	 		\caption{Various $k$ with $p = 5,000,p^*=71, n=100,000,X\sim t(2)$}\label{klep1}
\begin{tabular}{r|rrrrr}
 	 			\hline
 $k$ & Full & LEV($250$) & ALEV($250$)& SIS-IBOSS($250$)& UNIF\\  \hline
1,000& 1132.2454& 21.5109&  15.4830& 13.2781&    8.0193\\
2,000& 1132.2454& 32.9537&  29.4339& 25.5884& 25.7788\\
3,000& 1132.2454& 46.2533&  44.9549& 38.6782& 50.7871\\
4,000& 1132.2454& 68.6018&  68.0617& 61.3120&  96.7474\\
5,000& 1132.2454& 98.1117&  97.1221& 91.4450&  159.9517\\ \hline
 	 		\end{tabular}
 	 \end{subtable}
\end{table}

Table \ref{table3} compares the SIS-LASSO algorithm with model fitting using the full data, and the subdata approaches using the leverage sampling methods (denoted as LEV for exact leverage scores and ALEV for approximate leverage scores). As pointed out in \cite{drineas2012fast}, calculating exact leveraging scores with all variables could take longer time than fitting LASSO regression on the whole data set itself for large $p$. The approximate leverage sampling reduces computation time effectively when $n>>p$, but it is not applicable when $p$ becomes extremely large. (In our simulations we observed frequent computer crashes when using the leverage sampling methods for a large value of $p$.) To make the leverage sampling approach applicable to high-dimensional data sets, we proposed a SIS-type of adapted LEV and ALEV approaches. In Table \ref{table3}, we compare three approaches using the subset of $s$ variables having the largest absolute correlation with the response: SIS-IBOSS($s$), LEV($s$) and ALEV($s$).

  	 	 	\begin{table}[tb]
 	  	 		\centering
 	  	 			\caption{CPU Time (second) for SIS-IBOSS($s$) with different $s$, $p = 5000,p^*=71, X\sim t(2)$}
  \label{table4}
 	  	 		\begin{subtable}{0.5\textwidth}
 	  	 			\centering
 	  	 		\caption{Various $k$ with fixed $ n=10,000$}\label{deltavaries1}

 	  	 			\begin{tabular}{r|rrrr}
 	  	 				\hline
$k$ & $s=50$  & $s=$ 250  &$s=$ 500\\
 	  	 		\hline
1,000& 12.2658& 13.2781& 14.2985\\
2,000& 24.5311& 25.5884& 26.7924\\
3,000& 37.4174& 38.6782& 40.2774\\
4,000& 59.8233& 61.3120& 63.0818\\
5,000& 90.1589& 91.4450& 92.8520\\ \hline
 	  	 	\end{tabular}
 	  	 		\end{subtable}%
 	  	 		\begin{subtable}{0.5\textwidth}
 	  	 			\centering
 	  	 			\caption{Various $n$ with fixed $ k=1,000$}\label{deltavaries2}
 	  	 			\begin{tabular}{r|rrrr}
 	  	 				\hline
$n$ &$s=$ 50 &$s=$ 250  &$s=$ 500\\
 	  	 			\hline
10,000&  9.5412& 10.0679& 10.4890\\
20,000&  9.6662& 10.2062& 10.7415\\
40,000&  9.9117& 10.4938& 11.2467\\
80,000& 11.0417& 11.7900& 12.7519\\
100,000& 12.2658& 13.2781& 14.2985\\ \hline
 	  	 			\end{tabular}
 	  	 		\end{subtable}
 	  	 	\end{table}
 	  	 	  	 	 	
For SIS-IBOSS approach, it is important to choose a proper value of $\gamma$ (the proportion of variables that are selected in Step 1). As the partial sorting in IBOSS only takes a small amount of time compared to 10-fold cross-validation LASSO regression, it is expected to see only marginal increase in computation time when we increase $\gamma$. This is confirmed by the results shown in Table \ref{deltavaries1} and \ref{deltavaries2}. Consequently, the choice of $\gamma$ should be made based on optimizing estimation accuracy, which will be discussed in the next subsection.

Table \ref{spc1} compares SIS-IBOSS with the split-and-conquer and the uniform sampling approaches. In the table, SPC($w/K$) represents the split-and-conquer approach performed with parallel computation on $K$ computer cores, while SIS-IBOSS$(s)$ and UNIF do not use multi-cores. For the split-and-conquer approach, we recorded the CPU time that includes only the maximum time over the $K$ cores (instead of the sum of CPU times) and the cost of combining the coefficients. Even by this definition in favor of the split-and-conquer approach, the split-and-conquer approach took significant longer time compared to SIS-IBOSS and UNIF methods in all settings.
	\begin{table}
		\centering
		\caption{CPU Time (second) for various $n$ with $p = 5,000,p^*=71, k=1,000,X\sim t(2)$}\label{spc1}
		\begin{tabular}{r|rrrrr}
			\hline
$n$ &  UNIF&SIS-IBOSS(250)&SPC(2/5) &SPC(5/10) &SPC(10/20)  \\
\hline
10,000& 9.7813&  10.0679&  28.7090&  16.2194&  15.7249\\
20,000& 9.6441&  10.2062& 103.8247&  34.3779&  26.4275\\
40,000& 8.9245&  10.4938& 218.6580& 118.7682&  52.8553\\
80,000& 7.9334&  11.7900& 294.0590& 250.6395& 157.5920\\
100,000& 8.0193&  13.2781& 322.7556& 263.5747& 267.4637\\ \hline
		\end{tabular}
	\end{table}

\subsection{Prediction Accuracy} 	
To evaluate the prediction accuracy we used the mean squared error (MSE), which is calculated as
\begin{equation}\label{MSE}
MSE =\dfrac{1}{n_t} \sum_{i=1}^{n_t}\left(\bm{X}_t\bm{\beta}-\bm{X}_t\bm{\hat{\beta}}^{LASSO}\right)^2,
\end{equation}
where $n_t$ is the number of points in the test data set $\bm{X}_t$.
In all simulations, we set $n_t=1,000$. All the test data sets are generated from the same distribution as the one for the corresponding training data sets.
 	
\begin{figure}[t]
 		\caption{Test MSE for increasing $n$ with fixed $k=1000$, $p=500, p^*=23 $}\label{variesn}
 		\begin{subfigure}{0.5\textwidth}
 			\centering
 			\includegraphics[width=3.2in]{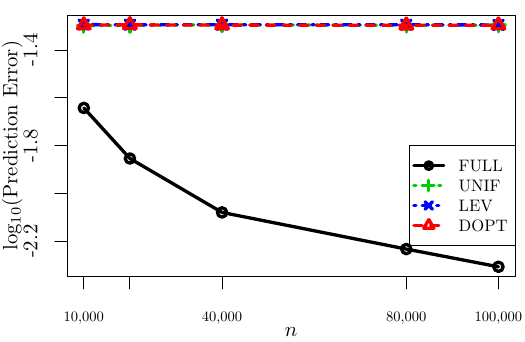}
 			\caption{$X\sim N(0,1)$}\label{errn1}
 		\end{subfigure}
 		\begin{subfigure}{0.5\textwidth}
 			\centering
 			\includegraphics[width=3.2in]{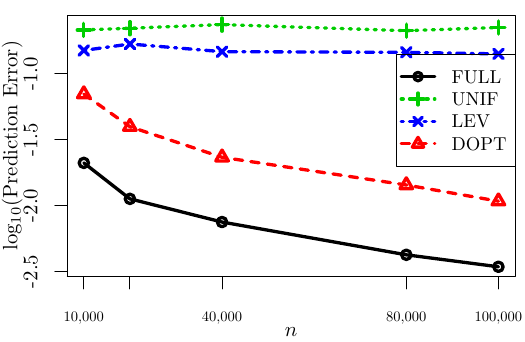}
 			\caption{	$X\sim t(2)$}\label{errt1}
 		\end{subfigure}
 		\begin{subfigure}{0.5\textwidth}
 			\centering
 			\includegraphics[width=3.2in]{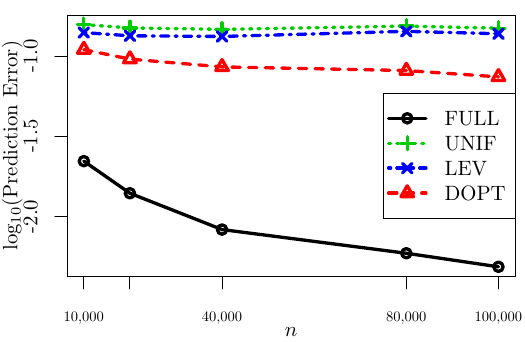}
 			\caption{$X\sim lognormal(0,1)$}\label{errln1}
 		\end{subfigure}
 		\begin{subfigure}{0.5\textwidth}
 			\centering
 			\includegraphics[width=3.2in]{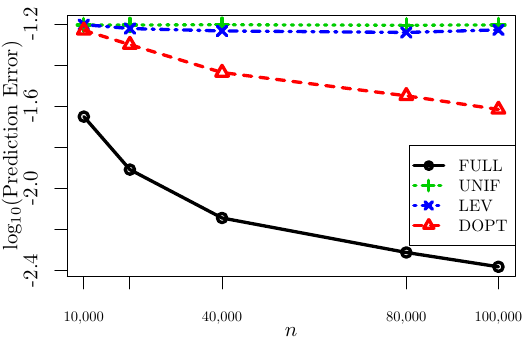}
 			\caption{	$X\sim Mixture$}\label{errm1}
 		\end{subfigure}
 	\end{figure}

\begin{figure}[!htb]
 		 	 		 		\caption{Test MSE for increasing $k$ with fixed $n=10^5, p=500,p^*=23$}\label{variesk}
 		\begin{subfigure}{0.5\textwidth}
 			\centering
 			\includegraphics[width=3.2in]{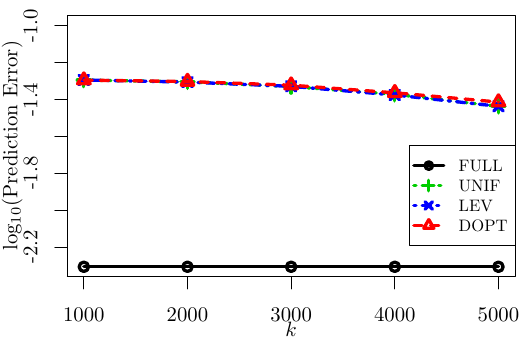}
 			\caption{$X\sim N(0,1)$}\label{errn2}
 		\end{subfigure}
 		\begin{subfigure}{0.5\textwidth}
 			\centering
 			\includegraphics[width=3.2in]{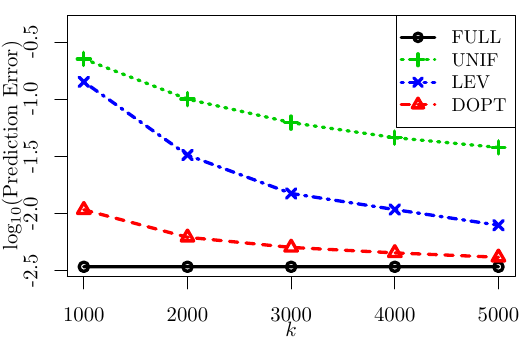}
 			\caption{	$X\sim t(2)$}\label{errt2}
 		\end{subfigure}
 		\begin{subfigure}{0.5\textwidth}
 			\centering
 			\includegraphics[width=3.2in]{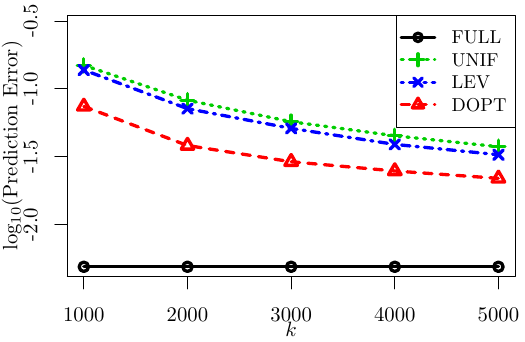}
 			\caption{$X\sim lognormal(0,1)$}\label{errln2}
 		\end{subfigure}
 		\begin{subfigure}{0.5\textwidth}
 			\centering
 			\includegraphics[width=3.2in]{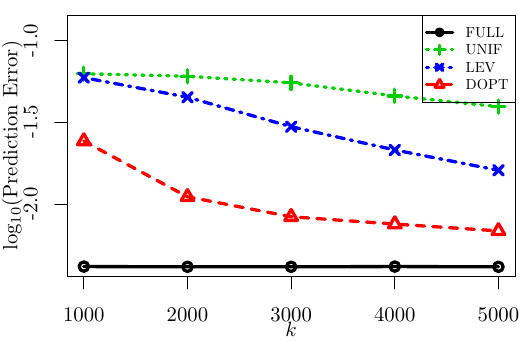}
 			\caption{	$X\sim Mixture$}\label{errm2}
 		\end{subfigure}
 	\end{figure}

Figures \ref{variesn} and \ref{variesk} display plots of $\log_{10}(MSE)$ values of several methods against various sizes of full data $n$ and sizes of subdata $k$, respectively. Compared to the LEV and UNIF methods, the IBOSS-LASSO approach (or, equivalently, $D$-optimal IBOSS) has smaller MSE in all settings of $n$ and $k$. This advantage is more significant when the distribution of $\bm{X}$ has heavier tails. As displayed in Figure \ref{variesn}, MSE from $D$-optimal IBOSS approach decreases when the full data size $n$ increases even though sample size is fixed at $k=1,000$. This is consistent with the theoretical property of IBOSS discussed in Wang et al. (2019). In comparison, the UNIF approach shows little change as $n$ increases. For LASSO regressions, MSE is also influenced by the strength of coefficient signals, but the relative difference of MSE across methods is fairly consistent no matter how large $|\beta_j|$ values are in our studies.

As expected, when LASSO was used on the full data set, the MSE decreases as $n$ increases. It is noteworthy that the performance of $D$-optimal IBOSS is comparable to some full data estimation. For example, in Figure \ref{errt1}, a $D$-optimal IBOSS subset of size $k=1,000$ from full data of size $100,000$ outperforms estimator by a full data of size 10,000, using only around $1/5$ of time (see Table \ref{tb2}). In Figure \ref{errt2}, $D$-optimal IBOSS estimator's MSE is less than $1/10$ of LEV estimator's when $k=10^3$.


 	
\begin{figure}[thbp]
 	 		\centering
\caption{Test MSE for $p=5,000$ for various $n$ and $k$}
\label{figure3}
 	 		 	\begin{subfigure}{0.5\textwidth}
 	 		 		\centering
 	 		 		\includegraphics[width=3.2in]{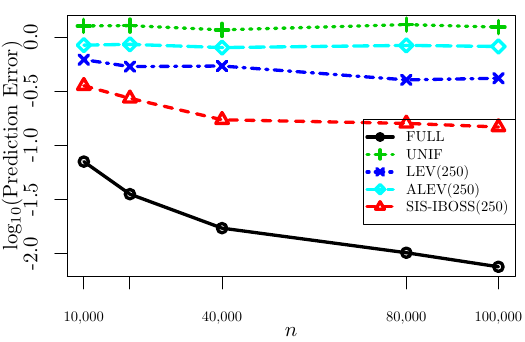}
 	 		 		\caption{	$X\sim t(2), k=1000$}\label{t5000nvaries}
 	 		 	\end{subfigure}%
 	 		 	\begin{subfigure}{0.5\textwidth}
 	 		 		\centering
 	 		 		\includegraphics[width=3.2in]{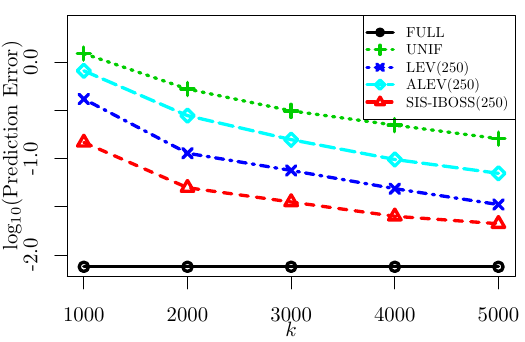}
 	 		 		\caption{	$X\sim t(2), n =10^5$}\label{t5000kvaries}
 	 		 	\end{subfigure}
 	 	\end{figure}

In Figure \ref{figure3}, we evaluated the performances of three SIS-enhanced methods (using $s=250$): SIS-LEV, SIS-ALEV, and SIS-IBOSS and compared them with models using the full data and the uniform sampling. The results were similar to those in Figures \ref{variesn} and \ref{variesk}. In almost all cases across various values of $n$ and $k$, SIS-IBOSS outperformed competing methods except the model using full data. As $k$ increased from 1,000 to 5,000, MSE of the SIS-IBOSS became very close to that of the full data model, as shown in Figure \ref{t5000kvaries}. The result again demonstrates the attractive feature of the proposed SIS-IBOSS approach as it used only 5\% of the total of 100,000 data points and took only 1/10 of the computation time of the corresponding full data model.


 	
\begin{figure}[htpb]
 	 	 	 		\centering
 	 	 	 		\caption{Test MSE for SIS-IBOSS with different $s = \gamma p$, $p=5,000$}
 \label{figure4}
 	 	 	 		\begin{subfigure}{0.5\textwidth}
 	 	 	 			\centering
 	 	 	 			\includegraphics[width=3.2in]{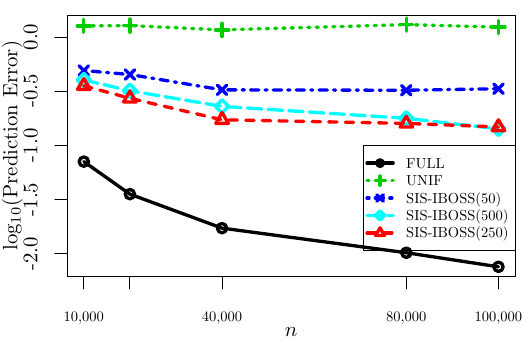}
 	 	 	 			\caption{	$X\sim t(2), k=1000$}\label{SISrvariesnvaries}
 	 	 	 		\end{subfigure}%
 	 	 	 		 	 	 	 		\begin{subfigure}{0.5\textwidth}
 	 	 	 		 	 	 	 			\centering
 	 	 	 		 	 	 	 			\includegraphics[width=3.2in]{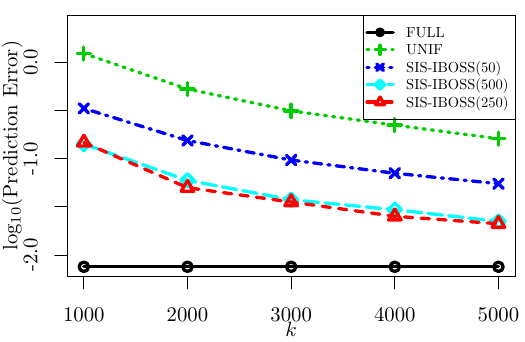}
 	 	 	 		 	 	 	 			\caption{	$ X\sim t(2),n= 10^5$}\label{SISrvarieskvaries}
 	 	 	 		 	 	 	 		\end{subfigure}
 	 	 	 	\end{figure}

We showed previously in Table \ref{table3} that the change in the proportion $\gamma$ had little impact on the computation time. Thus, $\gamma$ should be chosen to minimize the prediction error. Intuitively, if the true value of the number of significant variables $p^*$ were known, an ideal value of $\gamma$ could be chosen to be the smallest $\gamma$ satisfying $s=\gamma p > p^*$. In Figure \ref{figure4}, we compared three SIS-IBOSS methods for $s=50, 250,$ and 500, assuming $p^*=71$ and $p=5,000$. Results using the full data and the uniform sampling approach were also shown for benchmark purpose. The results are consistent with the intuition. In both cases, $s=250$ is a better choice than $s=50$, and increasing $s$ from 250 to 500 did not improve the performance much.

\begin{figure}[htpb]
 	 	 	 		\centering
 	 	 	 		\caption{MSE comparison with Split-and-Conquer approach}
 \label{figure5}

\begin{subfigure}{0.5\textwidth}
\centering
\includegraphics[width=3.2in]{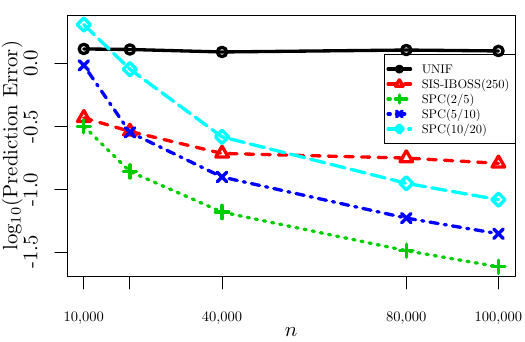}
\caption{Split and Conquer	$X\sim t(2)$}\label{spcnvaries}
\end{subfigure}
  	 	 		\begin{subfigure}{0.5\textwidth}
  	 	 			\centering
  	 	 			\includegraphics[width=3.2in]{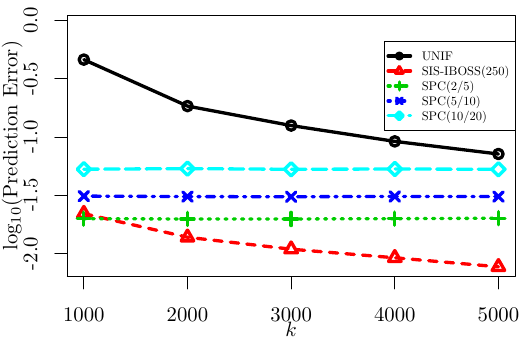}
  	 	 			\caption{Split and Conquer	$X\sim t(2)$}\label{spckvaries}
  	 	 		\end{subfigure}
  	 	 		
\end{figure}

We also compared our approaches with the split-and-conquer approach. In Figure \ref{figure5}, we consider studies where $p=5,000$ and $p^*=71$. Three different settings of the split-and-conquer approach are compared to SIS-IBOSS method, which include SPC(3/5) for $K=5,w=3$, SPC(5/10) for $K=10,w=5$, and SPC(10/20) for $K=20,w=10$. The split-and-conquer approach does not always beat other methods in terms of MSE, even though it takes longer computation time on $K$ cores (as shown previously in Table \ref{spc1}).

\subsection{Variable Selection Performance}
Variable selection is also an important perspective for LASSO type of models. Two commonly used criteria are {\em sensitivity} (true positive rate) and  {\em specificity} (true negative rate). Tables \ref{select_p5000_variesn} and \ref{select_p5000_variesk} compare sensitivity and specificity values of SIS-IBOSS with competing methods across various $n$ and $k$, respectively. In all cases, SIS-IBOSS(250) resulted in much higher sensitivity compared to UNIF, LEV(250), and ALEV(250) approaches. As a natural trade-off, the former method resulted in a smaller specificity, but the loss in specificity was minor and negligible in most cases. For example, Table \ref{select_p5000_variesn} shows that for $n=10,000$ and $k=1,000$, SIS-IBOSS(250) has the sensitivity of 0.9890, which is significantly higher than 0.9045 of LEV(250), the closest competitor except the model using the full data. In terms of specificity, SIS-IBOSS(250) has 0.9716, which is very close to 0.9729 of LEV(250).

We also compared three SIS-IBOSS($s$) for $s=50, 250$, and 500, when the true value of $p^*=71$. The results of Tables \ref{select_p5000_variesn} and \ref{select_p5000_variesk} further confirm that $s=250$ is superior to $s=50$ and comparable to $s=500$.

 \begin{table}[htpb]
 	\centering
 	\caption{$p=5,000$, $X_j\sim t(df=2),\epsilon_i\sim N(0,1) ,p^*=71$}\label{select_p5000_variesn}
 			\begin{subtable}{1.0\textwidth}
 				\centering
 				\caption{Sensitivity for increasing $n$ with a fixed $k=1,000$}\label{sensp5000nvaries}
 				\begin{tabular}{l|lllllll}
 					\hline
$n$ &FULL&      UNIF&      \multicolumn{3}{c}{SIS-IBOSS($s$)} &    LEV(250)&  ALEV(250)\\ \cline{4-6}
    &   &   & $s=50$ & $s=250$ & $s=500$    &   & \\ \hline
10,000&   0.9997&    0.6392&       0.8462&        0.9890&        0.9738&                0.9045&         0.8025\\
20,000&   1.0000&    0.6075&       0.8560&        0.9984&        0.9932&                 0.9377&         0.8237\\
40,000&   1.0000&    0.6315&       0.8646&        0.9993&        0.9869&                   0.8816&         0.8239\\
80,000&   1.0000&    0.6145&       0.8701&        0.9999&        0.9999&                 0.9360&         0.8013\\
100,000&   1.0000&    0.6243&       0.8675&        1.0000&        0.9998&               0.9353&         0.7848\\
 					\hline
 				\end{tabular}				
 			\end{subtable}	%
 			\begin{subtable}{1.0\textwidth}
 				\centering
 				\caption{Specificity for increasing $n$ with a fixed $k=1,000$}
 					\begin{tabular}{l|lllllll}
 						\hline
$n$ &FULL&      UNIF&      \multicolumn{3}{c}{SIS-IBOSS($s$)} &    LEV(250)&  ALEV(250)\\ \cline{4-6}
    &   &   & $s=50$ & $s=250$ & $s=500$    &   & \\ \hline
10,000&    0.9866 &   0.9879 &      0.9787 &       0.9716 &       0.9730 &               0.9729&         0.9793\\
20,000&    0.9902 &   0.9886 &      0.9790 &       0.9690 &       0.9717 &                0.9689&         0.9758\\
40,000&    0.9945 &   0.9884 &      0.9785 &       0.9687 &       0.9705 &            0.9724&         0.9764\\
80,000&    0.9972 &   0.9885 &      0.9784 &       0.9679 &       0.9688 &            0.9692&         0.9760\\
100,000&    0.9976 &   0.9891 &      0.9787 &       0.9677 &       0.9722 &             0.9701&         0.9760\\

 						\hline
 					\end{tabular}	\end{subtable}
 \end{table}

 	\begin{table}
 		\centering
 		 	\caption{$p=5,000$, $X_j\sim t(df=2),\epsilon_i\sim N(0,1) ,p^*=71$}\label{select_p5000_variesk}
 				\begin{subtable}{1.0\textwidth}
 					\centering
 					\caption{Sensitivity for increasing $k$ with a fixed $n=100,000$ and $p=5,000$}
 					\begin{tabular}{l|lllllll}
 						\hline
$k$ &FULL&      UNIF&      \multicolumn{3}{c}{SIS-IBOSS($s$)} &    LEV(250)&  ALEV(250)\\ \cline{4-6}
    &   &   & $s=50$ & $s=250$ & $s=500$    &   & \\ \hline
1,000& 1.0000   & 0.6243  &0.8675 &1.0000&  0.9998           & 0.9353       &  0.7848\\
2,000& 1.0000   & 0.9583  &0.9755 &1.0000&  1.0000          & 0.9986       &  0.9869\\
3,000& 1.0000   & 0.9943  &0.9959 &1.0000&  1.0000          & 0.9997       &  0.9994\\
4,000& 1.0000   & 0.9989  &0.9980 &1.0000&  1.0000            & 1.0000       &  1.0000\\
5,000& 1.0000   & 0.9997  &0.9993 &1.0000&  1.0000              & 1.0000       &  1.0000\\

 						\hline
 					\end{tabular}
 				\end{subtable}
 				\begin{subtable}{1.0\textwidth}
 					\centering
 					\caption{Specificity for increasing $k$ with a fixed $n=100,000$ and $p=5,000$}
 					\begin{tabular}{l|lllllll}
 						\hline
$k$ &FULL&      UNIF&      \multicolumn{3}{c}{SIS-IBOSS($s$)} &    LEV(250)&  ALEV(250)\\ \cline{4-6}
    &   &   & $s=50$ & $s=250$ & $s=500$    &   & \\ \hline
$1\times10^3$&0.9976 &0.9891 &0.9787  &0.9677  &0.9722  &0.9701  &0.9760\\
$2\times10^3$&0.9976 &0.9807 &0.9775  &0.9724  &0.9668  &0.9543  &0.9610\\
$3\times10^3$&0.9976 &0.9791 &0.9778  &0.9715  &0.9677  &0.9481  &0.9638\\
$4\times10^3$&0.9976 &0.9818 &0.9774  &0.9743  &0.9716 &0.9549  &0.9593\\
$5\times10^3$&0.9976 &0.9808 &0.9791  &0.9772  &0.9739  &0.9592  &0.9628\\

 						\hline
 					\end{tabular}
 				\end{subtable}
 	\end{table}

\section{Real Data Example}

Social media has become increasingly popular and important over the past few years. We considered real data
originated from blog posts (https://archive.ics.uci.edu/ml/datasets/BlogFeedback). The goal was to predict the number of comments to a blog post. More specifically, the response was the number of feedback comments of a blog within 24 hours the post. The raw HTML-documents
of the blog posts were crawled and processed. The data publishers chose a base time 
and selected the blog posts that were published no earlier than
72 hours before the selected base date/time. There were a total of 280 variables that were calculated by aggregating the selected blog posts from the information
that was available at the base time. Each observation
corresponded to a blog post.

There were 52,397 observations. To select the best predictors among the 280 variables, we fitted a linear LASSO model to a random subset of 50,000 observations and test the model with the 2,397 observations left.  In the data 280 variables included average, standard deviation, min, max and median of the length of time between the publication of the blog post and ``current'' time, the length of the blog post and so on. 
In LASSO regressions, $\lambda$ was selected by cross-validation with all the methods. For all the sub-sample approaches(D-OPT, LEV and UNIF), we used $k=5,600$ as sub-sample size. For each of the 280 variables, IBOSS selects 20 extreme points.


Since we do not know the true value of $\beta$, mean squared prediction error (MSPE) is utilized to compare the performances of different methods.  MSPE is defined as  
\begin{equation*}
MSPE =\dfrac{1}{n_t} \sum_{i=1}^{n_t}\left(y_t-\boldsymbol{X}_t\boldsymbol{\hat{\beta}}^{LASSO}\right)^2.
\end{equation*}
Notice that MSPE is larger than MSE defined in (5.1) since it contains the additional variance of the error term. The averaged test error over 1,000 runs of the
models for each method was reported in Table 8. While the size of the full data in the training set is only 50,000, the proposed IBOSS-LASSO approach resulted in a smaller MSPE value compared
to leverage sampling and uniform sampling methods, consistent with the findings in simulation studies.

\begin{table}
	\centering
	\caption{Prediction Error on Blog Feedback Dataset}\label{exampleerr}
	\begin{threeparttable}
		\begin{tabular}{l|lll}
			\hline
			FULL &D-OPT&LEV &UNIF\\\hline
			966.9707 &1105.0687& 1125.3645& 1129.3315 \\
			\hline
		\end{tabular}
	\end{threeparttable}
\end{table}

 	\section{Concluding Remarks}
In this article, we study the subdata selection under LASSO regression when both $n$ and $p$ are large. We establish the connection between $\bm{X}^T\bm{X}$ and the LASSO estimator $\hat{\beta}_n^{LASSO}$ when the tuning parameter $\lambda_n$ is assumed to grow slower than the square root of $n$. 
When $p$ is extremely large, a new algorithm combining SIS algorithm and IBOSS algorithm is proposed. Extensive numerical studies confirm some theoretical properties of both IBOSS-LASSO and SIS-IBOSS algorithms.
The numerical results also demonstrate that the proposed algorithms outperform the random sampling approach, in terms of both computational complexity and statistical efficiency.
The numerical results also show that the SIS-IBOSS algorithm sometimes can outperform the split-and-conquer approach in terms of both computational complexity and statistical efficiency.


There are important and unsolved questions that require future studies. First, the proposed algorithms depend on the model format. Due to the complexity of variable selection, we only consider linear models with main effects only in this manuscript. For many big data problems, however, a main-effect only model may not be adequate. For example, it has been recognized in many fields that second- or higher-order effects can be significant. Under a different model format, how to derive efficient algorithms for subdata selection in the context of variables selection would be an interesting and promising research field. Second, while a theoretical justification of the proposed algorithms is given, it is based on the LASSO asymptotic property. A direct justification would help us better understand the mechanism of subdata selection, thus develop more efficient algorithms. Unfortunately, such justification would require the explicit expression of the variance-covariance matrix of the LASSO estimator, which would be rather challenging because there exists no explicit expression of the LASSO estimator. 


\bibliographystyle{natbib}
\bibliography{ref}
 	
\end{document}